# Design and Simulation of an 8-bit Dedicated Processor for calculating the Sine and Cosine of an Angle using the CORDIC Algorithm

Aman Chadha[1,a], Divya Jyoti[2,b] and M. G. Bhatia[3,c]
[1,2] Thadomal Shahani Engineering College, Bandra (W), Mumbai, INDIA
[3] Ameya Centre for Robotics and Embedded Technology, Andheri (W), Mumbai, INDIA
[a] aman.x64@gmail.com, [b] dj.rajdev@gmail.com, [c] mgbhatia@acret.in

*Abstract* - **This paper describes the design and simulation of an 8-bit dedicated processor for calculating the Sine and Cosine of an Angle using CORDIC Algorithm (COordinate Rotation DIgital Computer), a simple and efficient algorithm to calculate hyperbolic and trigonometric functions. We have proposed a dedicated processor system, modeled by writing appropriate programs in VHDL, for calculating the Sine and Cosine of an angle. System simulation was carried out using ModelSim 6.3f and Xilinx ISE Design Suite 12.3. A maximum frequency of 81.353 MHz was reached with a minimum period of 12.292 ns. 126 (3%) slices were used. This paper attempts to survey the existing CORDIC algorithm with an eye towards implementation in Field Programmable Gate Arrays (FPGAs). A brief description of the theory behind the algorithm and the derivation of the Sine and Cosine of an angle using the CORDIC algorithm has been presented. The system can be implemented using Spartan3 XC3S400 with Xilinx ISE 12.3 and VHDL.**

## I. INTRODUCTION

Over the years, the field of Digital Signal Processing (DSP) has been essentially dominated by Microprocessors. This is mainly because of the fact that they provide designers with the advantages of single cycle multiply-accumulate instruction as well as special addressing modes [4]. Although these processors are cheap and flexible, they are relatively less time-efficient when it comes to performing certain resource-intensive signal processing tasks, e.g., Image Compression, Digital Communication and Video Processing. However as a direct consequence of rapid advancements in the field of VLSI and IC design, special purpose processors with custom-architectures are designed to perform certain specific tasks. They need fewer resources and are less complex than their general purpose counterparts. Instructions for performing a task are hardwired into the processor itself, i.e., the program is built right into the microprocessor circuit itself [2]. Due to this, the execution time of the program is considerably less than that if the instructions are stored in memory. Emerging high level hardware description and synthesis technologies in conjunction with Field Programmable Gate Arrays (FPGAs) have significantly lowered the threshold for hardware development as opportunities exist to integrate these technologies into a tool for exploring and evaluating micro-architectural designs [4]. Because of their advantage of real-time in-circuit reconfigurability, FPGAs based processors are flexible, programmable and reliable [1]. Thus, higher speeds can be achieved by these customized hardware solutions at competitive costs. Also, various simple and hardware-efficient algorithms exist which map well onto these chips and can be used to enhance speed and flexibility while performing the desired signal processing tasks [1],[2],[3].

One such simple and hardware-efficient algorithm is COordinate Rotation DIgital Computer (CORDIC) [5]. Primarily developed for real-time airborne computations, it uses a unique computing technique highly suitable for solving the trigonometric relationships involved in plane co-ordinate rotation and conversion from rectangular to polar form. John Walther extended the basic CORDIC theory to provide solution to and implement a diverse range of functions [7]. It comprises a special serial arithmetic unit having three shift registers, three adders/subtractors, Look-Up Table (LUT) and special interconnections. Using a prescribed sequence of conditional additions or subtractions, the CORDIC arithmetic unit can be designed to solve either of the following equations:

$$\begin{aligned} Y' &= K(Y\cos\lambda + X\sin\lambda) \\ X' &= K(X\cos\lambda - Y\sin\lambda) \end{aligned} \quad (1)$$

Where, K is a constant.

By making slight adjustments to the initial conditions and the LUT values, it can be used to efficiently implement trigonometric, hyperbolic, exponential functions, coordinate transformations etc. using the same hardware. Since it uses only shift-add arithmetic, the VLSI implementation of such an algorithm is easily achievable [4].

## II. CORDIC ALGORITHM

The CORDIC algorithm is an iterative technique based on the rotation of a vector which allows many transcendental and trigonometric functions to be calculated. The key aspect of this method is that it is achieved using only shifts, additions/subtractions and table look-ups which map well into hardware and are ideal for FPGA implementation. The CORDIC algorithms presented in this paper are well known in the research and super-computing circles.

*A. Algorithm Fundamentals*

Vector rotation is the first step to obtain the trigonometric functions. It can also be used for polar to rectangular and vice-versa conversions, for vector magnitude, and as a building block in certain transforms such as the Discrete Fourier Transform (DFT) and Discrete Cosine Transform (DCT). The algorithm is derived from Givens [6] rotation as follows:

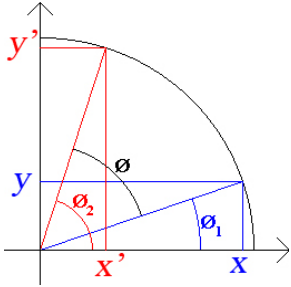

Fig. 1. Illustration of the CORDIC algorithm

In Fig. 1, the diagonal blue line is at an angle $\phi_1$ above the horizontal. The diagonal red line is actually the blue line rotated anti-clockwise by an angle $\phi$. The new X and Y values are related to the old X and Y values as follows:

$$x' = x\cos\phi - y\sin\phi$$
$$y' = y\cos\phi + x\sin\phi \quad (2)$$

For CORDIC, the final angle $\phi_2$ the angle whose sine or cosine we want to calculate and initial angle $\phi_1$ is set to a convenient value such as 0. Rather than rotating from $\phi_1$ to $\phi_2$ in one full sweep, we move in steps with careful choice of step values. Rearranging (2) gives us:

$$x' = \cos\phi \cdot [x - y\tan\phi]$$
$$y' = \cos\phi \cdot [y + x\tan\phi] \quad (3)$$

Restricting the rotation angles such that $\tan\phi = \pm 2^{-i}$, transforms the multiplication by the tangent term to a simple shift operation [1]. Arbitrary angles of rotation are obtained by successively performing smaller elementary rotations. If i, the decision at each iteration, is which direction to rotate rather than whether to rotate or not, then $\cos(\delta_i)$ is constant as $\cos(\delta_i) = \cos(-\delta_i)$. Then the iterative rotation can be expressed as:

$$x_{i+1} = K_i \left[ x_i - y_i \cdot d_i \cdot 2^{-i} \right]$$
$$y_{i+1} = K_i \left[ y_i + x_i \cdot d_i \cdot 2^{-i} \right] \quad (4)$$

Where, $K_i = \cos(\tan^{-1} 2^{-i}) = \dfrac{1}{\sqrt{1+2^{-i}}} = \left(\sqrt{1+2^{-i}}\right)^{-1}$

$$d_i = \pm 1$$

Removing the scale constant from the iterative equations yields a shift-add algorithm for vector rotation. The product of the $K_i$'s can be applied elsewhere in the system or treated as part of a system processing gain. That product approaches 0.6073 as the number of iterations reaches infinity. Therefore, the rotation algorithm has a gain, $A_n \approx 1.65$. The exact gain depends on the number of iterations, and follows the following equation:

$$A_n = \prod_n \sqrt{1 + 2^{-2i}} \quad (5)$$

The angle of a composite rotation is realized by the sequence of the directions of the elementary rotations. That sequence can be represented by a decision vector. The set of all possible decision vectors is an angular measurement system based on binary arctangents. Conversions between this angular system and any other can easily be accomplished using a LUT. A better conversion method uses an additional adder-subtractor that accumulates the elementary rotation angles post iteration. The angle accumulator adds a third difference equation to the CORDIC algorithm:

$$z_{i+1} = z_i - d_i \cdot \tan^{-1}(2^{-i}) \quad (6)$$

As discussed above, when the angle is in the arctangent base, this extra element is not needed. The CORDIC rotator is normally operated in one of two modes, i.e., the Rotation mode and the Vectoring mode.

*B. Rotation Mode*

The first mode of operation, called rotation by Volder [5],[4], rotates the input vector by a specified angle (given as an argument). Here, the angle accumulator is initialized with the desired rotation angle. The rotation decision based on the sign of the residual angle is made to diminish the magnitude of the residual angle in the angle accumulator. If the input angle is already expressed in the binary arctangent base, the angle accumulator is not needed [4],[1]. The equations for this are:

$$x_{i+1} = x_i - y_i \cdot d_i \cdot 2^{-i}$$
$$y_{i+1} = y_i + x_i \cdot d_i \cdot 2^{-i}$$
$$z_{i+1} = z_i - d_i \cdot \tan^{-1}(2^{-i}) \quad (7)$$

Where, $d_i = \begin{cases} -1 & \text{if } z_i < 0 \\ +1 & \text{otherwise} \end{cases}$

$$x_n = A_n[x_0 \cos z_0 - y_0 \sin z_0]$$
$$y_n = A_n[y_0 \cos z_0 + x_0 \sin z_0]$$
$$z_n = 0 \quad (8)$$
$$A_n = \prod_n \sqrt{1 + 2^{-2i}}$$

*C. Vectoring Mode*

In the vectoring mode, the CORDIC rotator rotates the input vector through whatever angle is necessary to align the result vector with the x axis. The result of the vectoring operation is a rotation angle and the scaled magnitude i.e. the x component of the original vector. The vectoring function works by seeking to minimize the y component of the residual vector at each rotation. The sign of the residual y component is used to determine which direction to rotate next. When initialized with zero, accumulator contains the traversed angle at the end of the iterations [4]. The equations in this mode are:

$$\begin{aligned} x_{i+1} &= x_i - y_i \cdot d_i \cdot 2^{-i} \\ y_{i+1} &= y_i + x_i \cdot d_i \cdot 2^{-i} \\ z_{i+1} &= z_i - d_i \cdot \tan^{-1}(2^{-i}) \end{aligned} \quad (9)$$

Where, $d_i = \begin{cases} +1 & \text{if } y_i < 0 \\ -1 & \text{otherwise} \end{cases}$

Then:

$$\begin{aligned} x_n &= A_n \sqrt{x_0^2 + y_0^2} \\ y_n &= 0 \\ z_n &= z_0 + \tan^{-1}\left(\frac{y_0}{x_0}\right) \\ A_n &= \prod_n \sqrt{1 + 2^{-2i}} \end{aligned} \quad (10)$$

The CORDIC rotation and vectoring algorithms as stated are limited to rotation angles between $-\pi/2$ and $\pi/2$. For composite rotation angles larger than $\pi/2$, an additional rotation is required [1]. Volder [4] describes an initial rotation of $\pm \pi/2$. This gives the correction iteration:

$$\begin{aligned} x' &= -d \cdot y \\ y' &= d \cdot x \\ z' &= z + d \cdot \frac{\pi}{2} \end{aligned} \quad (11)$$

Where, $d_i = \begin{cases} +1 & \text{if } y < 0 \\ -1 & \text{otherwise} \end{cases}$

There is no growth for this initial rotation. Alternatively, an initial rotation of either $\pi$ or 0 can be made, avoiding the reassignment of the x and y components to the rotor elements. Again, there is no growth due to the initial rotation:

$$\begin{aligned} x' &= d \cdot x \\ y' &= d \cdot y \\ z' &= \begin{cases} z & \text{if } d = 1 \\ z - \pi & \text{if } d = -1 \end{cases} \end{aligned} \quad (12)$$

Where, $d_i = \begin{cases} -1 & \text{if } x < 0 \\ +1 & \text{otherwise} \end{cases}$

Both reduction forms assume a modulo $2\pi$ representation of the input angle. The second reduction may be more convenient when wiring is restricted, as is often the case with FPGAs.

### D. Evaluation of Sine and Cosine using CORDIC

In rotational mode the sine and cosine of the input angle can be computed simultaneously. Setting the y component of the input vector to zero reduces the rotation mode result to:

$$\begin{aligned} x_n &= A_n \cdot x_0 \cos z_0 \\ y_n &= A_n \cdot x_0 \sin z_0 \end{aligned} \quad (13)$$

If $x_0$ is equal to $1/A_n$, the rotation produces the unscaled sine and cosine of the angle argument, $z_0$. Very often, the sine and cosine values modulate a magnitude value. Using other techniques (e.g., a LUT) requires a pair of multipliers to obtain the required modulation. The algorithm performs the multiply as part of the rotation operation, and therefore eliminates the need for a pair of explicit multipliers. The output of the CORDIC rotator is scaled by the rotator gain. If the gain is not acceptable, a single multiply by the reciprocal of the gain constant placed before the CORDIC rotator will yield unscaled results [1].

### E. Advantages

- Number of gates required in hardware implementation on an FPGA, are minimum. Thus, hardware complexity is greatly reduced compared to other processors such as DSP multipliers. Hence, it is relatively simple in design.
- Due to reduced hardware requirement, cost of a CORDIC hardware implementation is less as only shift registers, adders and look-up table (ROM) are required.
- Delay involved during processing is comparable to that of a division or square-rooting operation.
- No multiplication and only addition, subtraction and bit-shifting operation ensures simple VLSI implementation.
- Either if there is an absence of a hardware multiplier (e.g. microcontroller, microprocessor) or there is a necessity to optimize the number of logic gates (e.g. FPGA), CORDIC is the preferred choice [4].

### F. Applications

- The algorithm was basically developed to offer digital solutions to the problems of real-time navigation in B-58 bomber [5].
- This algorithm finds use in 8087 Math coprocessor, the HP-35 calculator [8], radar signal processors [8] and robotics.
- CORDIC algorithm has also been described for the calculation of DFT, DHT, Chirp Z-transforms, filtering, Singular value decomposition and solving linear systems [4].

- Most calculators, especially the ones built by Texas Instruments and Hewlett-Packard use CORDIC algorithm for calculation of transcendental functions.

### III. SYSTEM ARCHITECTURE

In this paper, the FPGA implementation of simple 8-bit dedicated processor for calculating the sine and cosine of an angle using CORDIC Algorithm is presented. The processor was implemented by using Xilinx ISE Design Suite 12.3 and VHDL. Fig. 2 shows functional block diagram of our 8-bit processor. It mainly consists of an 8-bit multiplexers, registers, arithmetic logic unit (ALU), tri-state buffer, comparator, and control unit. The logic circuit for dedicated microprocessor is divided into two parts: the datapath unit and control unit [9].

Input of two registers can be either from an external data input or from the output of ALU unit. Two control signals ln_X and ln_Y select which of two sources are to be loaded into registers. Two control signals XLoad and YLoad load a value into respective registers. Bottom multiplier determines the source of two operands of ALU. This allows the selection of one of the two subtraction operations X-Y or Y-X. A comparator unit is used to test condition of equal to or greater than and it accordingly generates status signals. Tristate buffer is used for outputting result from register X.

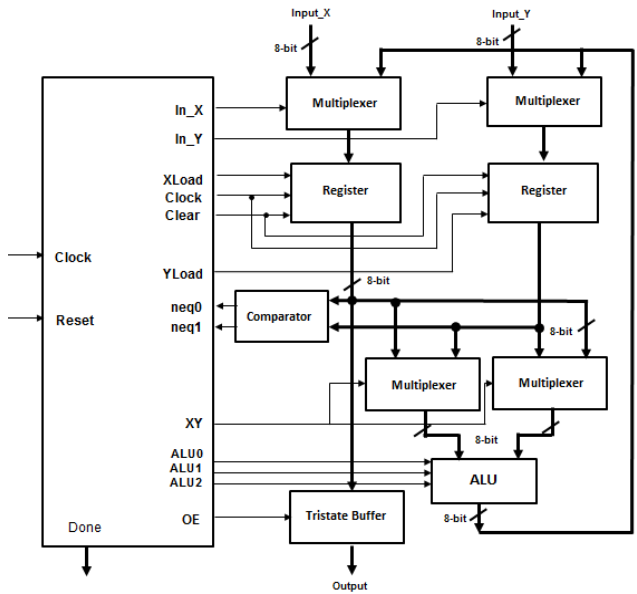

Fig. 2. Functional block diagram of the 8-bit processor

#### A. Datapath Unit

Datapath is responsible for the actual execution of all data operations performed by the dedicated processor [9]. Fig. 3 shows the datapath unit for the 8-bit dedicated processor.

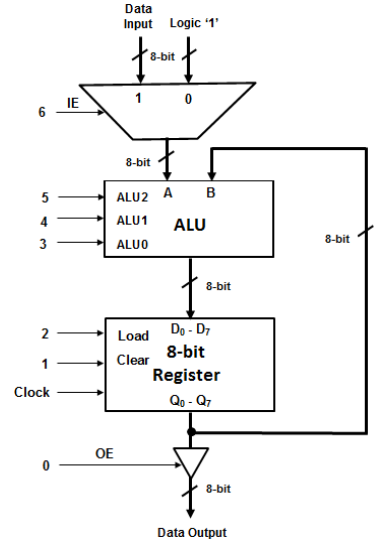

Fig. 3. A simple, general datapath circuit for the dedicated microprocessor

#### B. Control Unit

Fig. 4 shows the block diagram of control unit and Fig. 5 shows the corresponding state diagram.

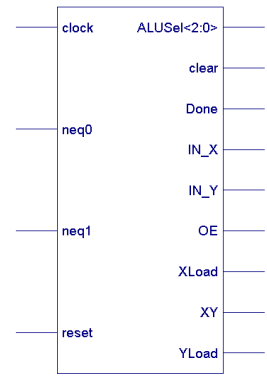

Fig. 3. Block diagram of the control unit

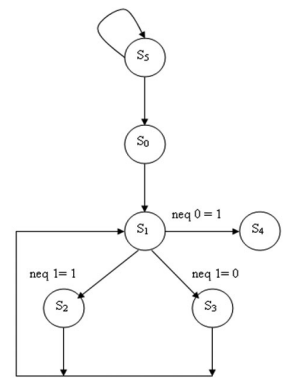

Fig. 4. State diagram of the control unit

Control signals are generated by the control unit which is modelled as a finite state machine with 6 states say $S_0$-$S_5$. There are 9 control signals which form control word and control the operation of datapath, as per the following table:

TABLE I
CONTROL SIGNAL STATUS DURING DIFFERENT STATES

| ln_X | ln_Y | XLoad | YLoad | XY | Clear |
|------|------|-------|-------|----|-------|
| 1 | 1 | 1 | 1 | 0 | 0 |
| 0 | 0 | 0 | 0 | 0 | 0 |
| 0 | 0 | 1 | 0 | 1 | 0 |
| 0 | 0 | 0 | 1 | 0 | 0 |
| 0 | 0 | 0 | 0 | 0 | 0 |
| 1 | 1 | 0 | 0 | 0 | 1 |

| ln_X | ln_Y | ALU(0,1,2) | OE | Done | State |
|------|------|------------|----|----|-------|
| 1 | 1 | 101 | 0 | 0 | $S_0$ |
| 0 | 0 | 101 | 0 | 0 | $S_1$ |
| 0 | 0 | 101 | 0 | 0 | $S_2$ |
| 0 | 0 | 101 | 0 | 0 | $S_3$ |
| 0 | 0 | 101 | 1 | 1 | $S_4$ |
| 1 | 1 | 101 | 0 | 0 | $S_5$ |

If Reset = '1' then state $S_5$ occurs. In this state registers are initialized to '0' by asserting the Clear signal. If Reset = '0' then at rising edge of clock, the state is upgraded from $S_5$ to state $S_0$. During the state $S_0$ two inputs are loaded in two registers. After completion of state $S_0$, state $S_1$ is reached. In this state output of registers is checked in comparator for equality and greater than conditions. If both values are same, state $S_4$ occurs else $S_2$ or $S_3$ will continue depending on status of signal neq1. State $S_1$ is repeated again. Default state is $S_5$.

## IV. IMPLEMENTATION AND VERIFICATION

All the units in dedicated processor were designed. These units were described in VHDL-modules and synthesized using ISE Design Suite 12.3. ModelSim simulator was used to verify the functionalities of each unit. Finally all the units were combined together and once again tested by using ModelSim simulator. Fig. 6 shows the RTL schematic of the CORDIC processor generated from Xilinx ISE.

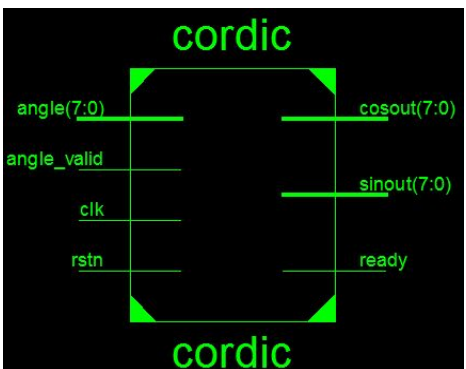

Fig. 6. RTL schematic of the CORDIC processor

### A. Datapath Unit

Simulation result of datapath is shown in Fig. 7.

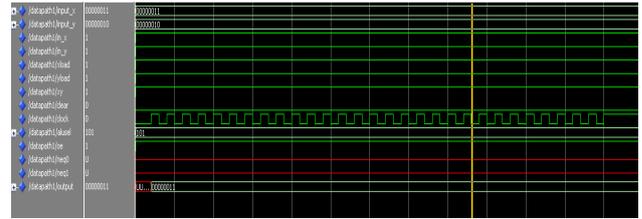

Fig. 7. Datapath unit simulation

The following table shows the synthesis report of the datapath unit:

TABLE II
SYNTHESIS REPORT OF THE DATAPATH UNIT

| Number of Slices | 61 (31%) |
|---|---|
| Maximum Frequency | 114.05 MHz |
| Minimum Period | 8.76 ns |

### B. Control Unit

Simulation result of the control unit is shown in Fig. 8.

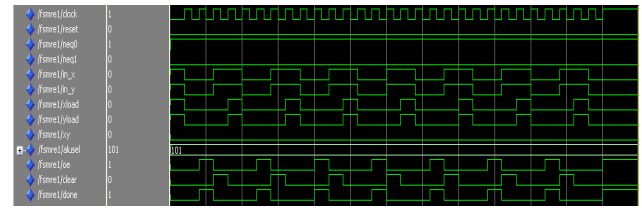

Fig. 8. Control unit simulation

The following table shows the synthesis report of the control unit:

TABLE III
SYNTHESIS REPORT OF THE CONTROL UNIT

| Number of Slices | 4 (2%) |
|---|---|
| Maximum Frequency | 264.34 MHz |
| Minimum Period | 3.78 ns |

### C. Dedicated CORDIC Processor

Once the datapath unit and control unit were simulated, they were combined and a dedicated processor was constructed. Simulation shows the CORDIC calculation operation of the Sine and Cosine of an angle. Simulation result for the dedicated processor is as shown in Fig. 9.

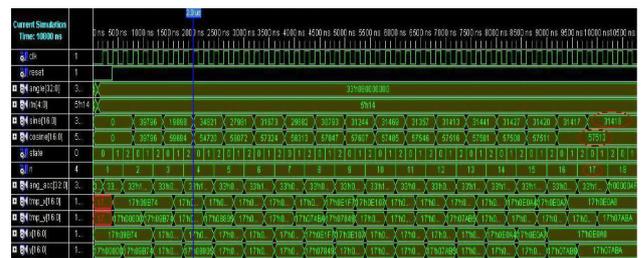

Fig. 9. Test-Bench waveforms indicating the Sine and Cosine output, obtained after the CORDIC Core Simulation

The following table shows the synthesis report of the 8-bit dedicated processor:

TABLE IV
SYNTHESIS REPORT OF THE 8-BIT DEDICATED PROCESSOR

| Logic Utilization | Used | Available | Utilization |
|---|---|---|---|
| Number of Slices | 126 | 3584 | 3% |
| Number of Slice Flip Flops | 58 | 7168 | 0% |
| Number of 4 input LUTs | 238 | 7168 | 3% |
| Number of bonded IOBs | 28 | 141 | 19% |
| Number of GCLKs | 1 | 8 | 12% |

For a Speed Grade of -5, the minimum period required is 12.292 ns, which corresponds to a maximum frequency of 81.353 MHz. The minimum input arrival time before clock and maximum output required time after clock are 9.657 ns and 8.133 ns respectively.

The following table shows the comparison between the actual Sine and Cosine values and the ones obtained from Test-Bench analysis in VHDL.

TABLE V
COMPARISON OF SUCCESSIVE ANGLE ROTATION VALUES

| Angle (A) | Rotation | sin(A) (Actual) | sin(A) (Test-Bench) | Error |
|---|---|---|---|---|
| 0.000000 | 5 | 0.00000000 | 0.01483516 | -1.4835e-002 |
| 0.000000 | 10 | 0.00000000 | 0.00117259 | -1.1725e-003 |
| 0.000000 | 15 | 0.00000000 | 0.00001292 | -1.2922e-005 |
| 0.000000 | 20 | 0.00000000 | -0.00000043 | 4.2874e-007 |
| 0.523599 | 5 | 0.50000000 | 0.48362630 | 1.6373e-002 |
| 0.523599 | 10 | 0.50000000 | 0.49892865 | 1.0713e-003 |
| 0.523599 | 15 | 0.50000000 | 0.50003905 | -3.9047e-005 |
| 0.523599 | 20 | 0.50000000 | 0.50000106 | -1.0561e-006 |
| 1.000000 | 5 | 0.84147098 | 0.80881306 | 3.2657e-002 |
| 1.000000 | 10 | 0.84147098 | 0.84080033 | 6.7065e-004 |
| 1.000000 | 15 | 0.84147098 | 0.84149350 | -2.2515e-005 |
| 1.000000 | 20 | 0.84147098 | 0.84147186 | -8.7478e-007 |
| 3.141593 | 5 | 0.00000000 | -0.01483516 | 1.4835e-002 |
| 3.141593 | 10 | 0.00000000 | -0.00117259 | 1.1725e-003 |
| 3.141593 | 15 | 0.00000000 | -0.00001292 | 1.2922e-005 |
| 3.141593 | 20 | 0.00000000 | 0.00000043 | -4.2874e-007 |

| Angle (A) | Rotation | cos(A) (Actual) | cos(A) (Test-Bench) | Error |
|---|---|---|---|---|
| 0.000000 | 5 | 1.00000000 | 0.99988995 | 1.1004e-004 |
| 0.000000 | 10 | 1.00000000 | 0.99999931 | 6.8748e-007 |
| 0.000000 | 15 | 1.00000000 | 1.00000000 | 8.3498e-011 |
| 0.000000 | 20 | 1.00000000 | 1.00000000 | 9.2037e-014 |
| 0.523599 | 5 | 0.86602540 | 0.87527459 | -9.2491e-003 |
| 0.523599 | 10 | 0.86602540 | 0.86664307 | -6.1766e-004 |
| 0.523599 | 15 | 0.86602540 | 0.86600286 | 2.2545e-005 |
| 0.523599 | 20 | 0.86602540 | 0.86602479 | 6.0975e-007 |
| 1.000000 | 5 | 0.54030231 | 0.58806584 | -4.7763e-002 |
| 1.000000 | 10 | 0.54030231 | 0.54134537 | -1.0430e-003 |
| 1.000000 | 15 | 0.54030231 | 0.54026724 | 3.5067e-005 |
| 1.000000 | 20 | 0.54030231 | 0.54030094 | 1.3623e-006 |
| 3.141593 | 5 | -1.0000000 | -0.99988995 | -1.1004e-004 |
| 3.141593 | 10 | -1.0000000 | -0.99999931 | -6.8748e-007 |
| 3.141593 | 15 | -1.0000000 | -1.00000000 | -8.3498e-011 |
| 3.141593 | 20 | -1.00000000 | -1.00000000 | -9.2037e-014 |

## V. CONCLUSION

We have successfully simulated an 8-bit dedicated processor for calculating the Sine and Cosine of an angle, on ModelSim simulator using the VHDL language. Our processor has six main components namely, control unit, multiplexer unit, ALU unit, register unit, tristate buffer unit and comparator. Our dedicated processor has a maximum frequency of 81.353 MHz was reached with a minimum period of 12.292 ns. 126 (3%) slices were used. Our System can be implemented on Xilinx Spartan 3 XC3S400 using ISE Design Suite 12.3 and VHDL language. Our dedicated processor has a distinct advantage over a general purpose processor, since it repeatedly performs same task its design is more efficient and consumes less resources and is less time intensive.